\documentclass[11pt]{article}

\usepackage{amssymb}
\usepackage{amsmath,amsthm}  

\usepackage{ulem} 

\usepackage{graphicx}

\usepackage[english,francais]{babel}

\newfont{\ensmathquatorze}{msbm10 scaled 1400}
\newfont{\ensmathonze}{msbm10 scaled 1100}
\newfont{\ensmathdix}{msbm10}
\newfont{\ensmathneuf}{msbm10 scaled 833}
\newfont{\ensmathhuit}{msbm10 scaled 694}
\newfam\ensmathfam                        
\textfont\ensmathfam=\ensmathonze        
\scriptfont\ensmathfam=\ensmathdix       
\scriptscriptfont\ensmathfam=\ensmathhuit
\def\ensmf{\fam\ensmathfam\ensmathonze}         

\newcommand{\ket}[1]{|\kern.3ex#1\kern.3ex\rangle}
\newcommand{\bra}[1]{\langle\kern.3ex #1 \kern.3ex|}

\newcommand{\EXP}[1]{{\mbox{\large e}}^{#1}}         

\newcommand{\cotg}{\mathop{\mathrm{cotg}}\nolimits}  

\def\NN{{\ensmf N}}                 
\def\RR{{\ensmf R}}                 

\def\I{{\rm i}}                  
\def\D{{\rm d}}                  
\def\Dc{{\rm D}}                 

\newcommand{\drond}[2]{\frac{\partial #1}{\partial #2}} 

\newcommand\ab{{\alpha\beta}}
\newcommand\ba{{\beta\alpha}}
\newcommand\mn{{\mu\nu}}
\newcommand\nm{{\nu\mu}}

\newcommand{\diagram}[3]{\raisebox{#3}{\includegraphics[scale=#2]{#1}}}

\newcommand\antiddots{\mathinner{\mkern2mu\raise1pt\hbox{.}\mkern2mu
\newline \raise4pt\hbox{.}\mkern2mu\raise7pt\hbox{.}\mkern1mu}}

\begin{document}

\selectlanguage{english}

\title{On the spectrum of the Laplace operator of metric graphs \\
       attached at a vertex -- Spectral determinant approach}

\author{Christophe Texier}
 
\date{January 31, 2008}

\maketitle  

\hspace{1cm}
\begin{minipage}[t]{13cm}
{\small

Laboratoire de Physique Th\'eorique et Mod\`eles Statistiques,
UMR 8626 du CNRS, 

Universit\'e Paris-Sud, B\^at. 100, F-91405 Orsay Cedex, France.

\vspace{0.25cm}

Laboratoire de Physique des Solides, 
UMR 8502 du CNRS, 

Universit\'e Paris-Sud, B\^at. 510, F-91405 Orsay Cedex, France.

}
\end{minipage}

\begin{abstract}
  We consider a metric graph $\mathcal{G}$ made of two graphs $\mathcal{G}_1$
  and $\mathcal{G}_2$ attached at one point. We derive a formula relating the
  spectral determinant of the Laplace operator
  $S_\mathcal{G}(\gamma)=\det(\gamma-\Delta)$ in terms of the spectral
  determinants of the two subgraphs. The result is generalized to describe the
  attachment of $n$ graphs. The formulae are also valid for the spectral
  determinant of the Schr\"odinger operator~$\det(\gamma-\Delta+V(x))$.
\end{abstract}

\noindent
PACS numbers~: 02.70.Hm~; 02.10.Ox 






\vspace{0.5cm}

\noindent{\bf Introduction.--}
Let us consider a bounded compact domain $\mathcal{D}_1$, part of a manifold.
We denote by $\mathrm{Spec}(-\Delta;\mathcal{D}_1)$ the set of solutions $E$
of $-\Delta\psi(r)=E\psi(r)$ with $\psi(r)$ satisfying given boundary
conditions at the boundary $\partial\mathcal{D}_1$ (Sturm-Liouville problem).
Similarly we consider a second bounded compact domain $\mathcal{D}_2$,
distinct from $\mathcal{D}_1$ and denote
$\mathrm{Spec}(-\Delta;\mathcal{D}_2)$ the spectrum of the Laplace operator in
$\mathcal{D}_2$. Now, if we can glue $\mathcal{D}_1$ and $\mathcal{D}_2$ by
identification of parts of $\partial\mathcal{D}_1$ and $\partial\mathcal{D}_2$
in order to form a unique compact domain $\mathcal{D}$, the question is~:  can
we relate the spectrum $\mathrm{Spec}(-\Delta;\mathcal{D})$ to
$\mathrm{Spec}(-\Delta;\mathcal{D}_1)$ and
$\mathrm{Spec}(-\Delta;\mathcal{D}_2)$~? The aim of this article is to 
discuss this question in the particular case of metric graphs when two
graphs are attached at one point. For that purpose the spectral information is
encoded in the spectral determinant of the graph $\mathcal{G}$, formally
defined as $S_\mathcal{G}(\gamma)=\det(\gamma-\Delta)$. We first define basic
notations and briefly recall some results on the spectral determinant of
metric graphs. We derive the relation between the spectral determinant of a
graph in terms of the two subgraph determinants, when subgraphs are attached
by one point, as represented on figure~\ref{fig:attachgraphs}.c. The relation
is generalized to describe attachment of $n>2$ graphs
(figure~\ref{fig:attachgraphs}.d) and to deal with Schr\"odinger operator.
It is interesting to point out that our result is reminiscent of the gluing
formula for elliptic operators acting on a manifold obtained in
Ref.~\cite{BurFriKap92}.

\vspace{0.25cm}

\noindent{\bf Metric graphs.--}
Let us consider a collection of $V$ vertices, denoted here by greek letters
$\alpha,\,\beta\,...$, connected between each others by $B$ bonds, denoted
$(\ab),\,(\mn)...$ Each bond is associated with two oriented bonds, that we
call arcs and denote as $\ab,\,\ba,\,\mn,\nm...$. The topology of the graph is
characterized by its adjacency (or connectivity) matrix $a_\ab$~:  $a_\ab=1$
if $(\ab)$ is a bond and $a_\ab=0$ otherwise. The coordination number of the
vertex $\alpha$ is denoted $m_\alpha=\sum_\beta a_\ab$. Up to now we have
built a ``combinatorial graph''. If now each bond is identified with an
interval $[0,l_\ab]\in\RR$, where $l_\ab$ is the length of the bond $(\ab)$,
the set of all connected bonds forms a ``metric graph'' (also called a
``quantum graph'').

A scalar function $\varphi(x)$ living on a graph $\mathcal{G}$ is defined by
$B$ components $\varphi_\ab(x_\ab)$ where $x_\ab\in[0,l_\ab]$ is the
coordinate along the bond ($x_\ab=0$ corresponds to vertex $\alpha$ and
$x_\ab=l_\ab$ to vertex $\beta$). By construction $x_\ab+x_\ba=l_\ab$. Note
that components are labelled by arc variables, since we must specify the
orientation of the axis along which the coordinate is given~; Obviously
$\varphi_\ab(x_\ab)=\varphi_\ba(x_\ba)$ for a scalar function. The action of
the Laplace operator on the scalar function along a bond coincides with the
one-dimensional Laplace operator $(\Delta\varphi)_\ab(x)=\varphi_\ab''(x)$. In
order to define a self-adjoint operator, one must specify boundary conditions
at the vertices. The most general conditions have been discussed in
Ref.~\cite{KosSch99}~(in general the question of boundary conditions is
related to the precise nature of the scattering at the vertex
\cite{GerPav88,ExnSeb89,Ada92,TexMon01}). In the present article we consider the
simple case of the Laplace operator acting on scalar functions that are
continuous at the vertices~:
$\varphi_\ab(x_\ab=0)=\varphi_\alpha\:\forall\:\beta$ neighbour of $\alpha$
(that gives $m_\alpha-1$ equations at the vertex $\alpha$ of coordination number
$m_\alpha$). Then one must impose another condition on derivatives of the
function. For continuous boundary condition, the most general condition that
ensures self-adjointness of Laplace operator is
$\sum_\beta{a}_\ab\varphi_\ab'(x_\ab=0)=\lambda_\alpha\,\varphi_\alpha$ with
$\lambda_\alpha\in\RR$. 
The presence of the adjacency matrix in the
sum, contraints this latter to run over vertices neighbour of $\alpha$ only.
The $m_\alpha$ equations ensure self adjointness of Laplace operator.
$\lambda_\alpha=\infty$ corresponds to Dirichlet boundary
condition~($\varphi_\alpha=0$).
%
The study of the Laplace operator on a metric graph appears in several
contexts, reviewed in Refs.~\cite{AkkComDesMonTex00,ComDesTex05}, like quantum
mechanical problems~: $-\Delta\varphi(x)=E\varphi(x)$ could be the Schr\"odinger
equation. In such a case it can be more interesting to consider the situation
of a graph submitted to a magnetic field, what is achieved by replacing the
derivative by the covariant derivative,
$\frac{\D}{\D{x}}\to\Dc_x=\frac{\D}{\D{x}}-\I{A}(x)$, where $A(x)$ is the
vector potential. The boundary condition then reads 
$\sum_\beta{a}_\ab(\Dc_x\varphi)_\ab(0)=\lambda_\alpha\,\varphi_\alpha$.

\begin{figure}[!ht]
  \centering
  \diagram{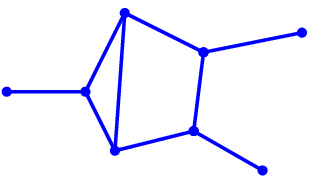}{1}{0cm}
  \hspace{2cm}
  \diagram{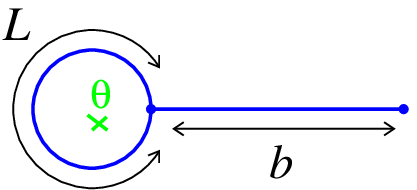}{1}{0cm}
  \caption{\it Examples. {\sf Left}~: a graph with $V=8$ vertices and $B=9$ 
           bonds. {\sf Right}~: a ring pierced by a magnetic flux 
           $\theta$ attached to a wire ($B=V=2$).}
  \label{fig:graphexample}
\end{figure}

\vspace{0.25cm}

\noindent{\bf Spectral determinant.--}
The spectral determinant of the Laplace operator $\Delta$ is formally defined
as $S_\mathcal{G}(\gamma)=\det(\gamma-\Delta)$, where $\gamma$ is the spectral
parameter. This object has been introduced in Ref.~\cite{PasMon99} in order to
study magnetization of networks of metallic wires. Despite the Laplace
operator acts in a space of infinite dimension, $S_\mathcal{G}(\gamma)$ can be
related to the determinant of a finite size matrix~\cite{PasMon99}~:
\begin{equation}
  \label{spedet}
  S_\mathcal{G}(\gamma) = 
  \prod_{(\ab)}\frac{\sinh\sqrt\gamma l_\ab}{\sqrt\gamma}\:
  \det\mathcal{M}
\end{equation}
Note that the first product, runing over all bonds, coincides with the
Dirichlet determinant (spectral determinant for Dirichlet conditions at all
vertices). A similar decomposition was obtained for combinatorial graphs in
Ref.~\cite{Col99}. The $V\times{V}$-matrix $\mathcal{M}$ has matrix elements~:
\begin{equation}
  \label{defmatM}
  \mathcal{M}_\ab = 
  \delta_\ab
  \left(\lambda_\alpha +\sqrt{\gamma} \sum_\mu a_{\alpha\mu}
  \coth\sqrt{\gamma}l_{\alpha\mu}\right)
  -a_\ab \frac{\sqrt\gamma\:\EXP{-\I\theta_\ab}}{\sinh\sqrt{\gamma}l_\ab}
  \:.
\end{equation}
This expression describes the case with magnetic field~:  $\theta_\ab$ is the
circulation of the vector potential along the wire
$\theta_\ab=\int_\ab\D{x}\,A(x)$. Generalization to the case of the spectral
determinant of the Schr\"odinger (Hill) operator
$S_\mathcal{G}(\gamma)=\det(\gamma-\Delta+V(x))$ with generalized boundary
conditions has been obtained in Refs.~\cite{Des00,Des01} (see also the
review articles~\cite{AkkComDesMonTex00,ComDesTex05}).
The result (\ref{spedet}) has been derived by two methods~: ({\it i})
construction and integration of the Kernel of the operator
$(-\Delta+\gamma)^{-1}$~\cite{PasMon99,AkkComDesMonTex00,Des00,Des01}~: 
$
\int_\mathcal{G}\D{x}\,\bra{x}\frac1{-\Delta+\gamma}\ket{x}
=\partial_\gamma\ln{S_\mathcal{G}(\gamma)}
$.
({\it ii}) A path integral derivation~\cite{AkkComDesMonTex00}. These
derivations give the spectral determinant, up to a numerical factor
independent on $\gamma$ (this is inessential for physical quantities since
they are always related to $\partial_\gamma\ln{S_\mathcal{G}}$). In the
present article, the precise prefactor of the spectral determinant is fixed by
eq.~(\ref{spedet}). Doing so we do not provide a way to determine the
$\gamma$-independent prefactor from the spectrum of the graph.

It is worth mentioning that the derivation of a $\zeta$-regularized
determinant allows to define the prefactor of the spectral determinant within
the calculation. If we denote $\{E_n\}$ the spectrum of an operator
$\mathcal{O}$, the determinant of this latter is defined thanks to the
$\zeta$-function $\zeta(s)=\sum_nE_n^{-s}$ as
$\det_\zeta\mathcal{O}=\exp-\zeta'(0)$~\cite{For87}. This approach has been
used in Ref.~\cite{Fri06} where result of Ref.~\cite{Des00} for continuous boundary
conditions has been obtained with a procedure fixing precisely the
multiplicative factor\footnote{
  We connect notations of Ref.~\cite{Fri06} with ours.
  Matrix $A\to$parameters $\lambda_\alpha$'s~;
  $\det(R(\lambda)+A)\to\det\mathcal{M}$~;
  the Dirichlet determinant is 
  $\det(H_D+\lambda)\to\prod_{(\ab)}\frac{2\sinh\sqrt\gamma l_\ab}{\sqrt\gamma}$.
  Therefore Eq.~(1.1) of Ref.~\cite{Fri06} for the $\zeta$-regularized
  spectral determinant shows that this latter is related to eq.~(\ref{spedet})
  by~$S^\zeta_\mathcal{G}(\gamma)=\frac{2^B}{\prod_\alpha{m_\alpha}}S_\mathcal{G}(\gamma)$. 
}.

\vspace{0.25cm}

\noindent{\bf Attachment of two graphs.--}
Let us consider two graphs $\mathcal{G}_1$ and $\mathcal{G}_2$ characterized
by matrices $\mathcal{M}_1^{\lambda_\alpha}$ and
$\mathcal{M}_2^{\lambda_\beta}$ for generalized boundary conditions at
vertices $\alpha$ and $\beta$, characterized by parameters $\lambda_\alpha$
and~$\lambda_\beta$. We denote by $S^{\lambda_\alpha}_1(\gamma)$ and
$S^{\lambda_\beta}_2(\gamma)$ the corresponding spectral determinants.

\newcommand{\taille}{0.7}
\begin{figure}[!ht]
  \centering
  \diagram{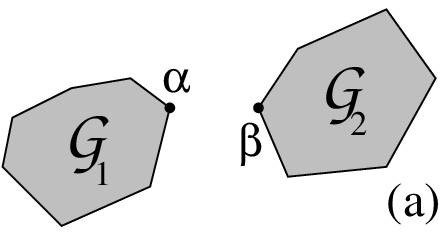}{\taille}{-1cm}
  \hspace{0.25cm}{\large$\longrightarrow$}\hspace{0.25cm}
  \diagram{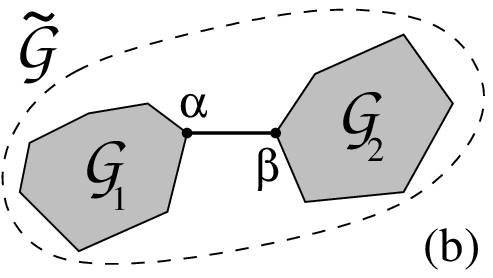}{\taille}{-1cm}
  \hspace{0.25cm}{\large$\longrightarrow$}\hspace{0.25cm}
  \diagram{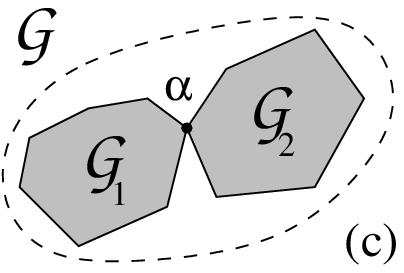}{\taille}{-1cm}
  \hspace{0.75cm}
  \diagram{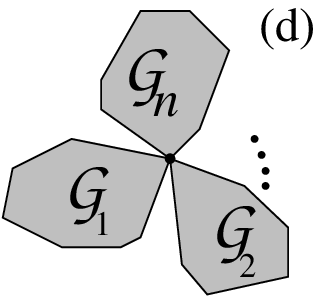}{\taille}{-1cm}
  \caption{\it (a) to (c)~: Attachment of two graphs $\mathcal{G}_1$ and $\mathcal{G}_2$
    (the dashed areas hide the structures of the graphs)~: 
    a bond $(\ab)$ is introduced, then he limit $l_\ab\to0$ is taken.}
  \label{fig:attachgraphs}
\end{figure}

We now attach $\mathcal{G}_1$ and $\mathcal{G}_2$ with a bond $(\ab)$
(figure~\ref{fig:attachgraphs}.b). The new graph is denoted
$\widetilde{\mathcal{G}}$. The matrix $\mathcal{M}$ characterizing the new
graph has the structure~:
\begin{equation}
  \label{structureM}
  \mathcal{M} = 
  \left(
    \begin{array}{ccc|ccc}
      & & & \vdots & & \antiddots \\
      & \mathcal{M}_1^{\lambda'_\alpha} & & 0 & 0 & \\
      & & & -\frac{\sqrt{\gamma}}{\sinh\sqrt{\gamma}l_\ab} & 0 & \cdots \\
      \hline
      \cdots & 0 & -\frac{\sqrt{\gamma}}{\sinh\sqrt{\gamma}l_\ab} & & & \\
      & 0 & 0 & & \mathcal{M}_2^{\lambda'_\beta} & \\
      \antiddots & & \vdots & & &
    \end{array}
  \right)
  \hspace{0.5cm}
\end{equation}
The diagonal blocks coincide with the matrices $\mathcal{M}_1^{\lambda_\alpha}$ and
$\mathcal{M}_2^{\lambda_\beta}$ of the isolated graphs, provided a
modification of the parameters describing boundary condition~: 
$\lambda_\alpha'=\lambda_\alpha+\sqrt{\gamma}\coth\sqrt{\gamma}l_\ab$ 
and $\lambda_\beta'=\lambda_\beta+\sqrt{\gamma}\coth\sqrt{\gamma}l_\ab$ 
(the coth accounts for the additional wire). Then we see that
\begin{equation}
  \det\mathcal{M} = 
  \det\left[\mathcal{M}_1^{\lambda'_\alpha}\right]\,
  \det\left[
    \mathcal{M}_2^{\lambda'_\beta}
    -\mathcal{J}^{(\beta)}\:\frac{\gamma}{\sinh^2\sqrt{\gamma}l_\ab}
     \left[(\mathcal{M}_1^{\lambda'_\alpha})^{-1}\right]_{\alpha\alpha}
  \right]
\end{equation}
where $\mathcal{J}^{(\beta)}$ is the matrix with only one non zero matrix element
equal to 1 on the diagonal corresponding to vertex $\beta$~:
$\mathcal{J}^{(\beta)}_\mn=\delta_{\mu\beta}\delta_{\nu\beta}$.
Below we introduce the notation 
$\mathcal{M}_1\equiv\mathcal{M}_1^{\lambda'_\alpha=\lambda_\alpha}$ and 
$\mathcal{M}_2\equiv\mathcal{M}_2^{\lambda'_\beta=\lambda_\beta}$ that denote
matrices characterizing $\mathcal{G}_1$ and $\mathcal{G}_2$ before connection.
Using
\begin{equation}
  \left(\left[\mathcal{M}_1^{\lambda'_\alpha}\right]^{-1}\right)_{\alpha\alpha}
=\left(
  \left[
    \mathcal{M}_1+\mathcal{J}^{(\alpha)}\sqrt{\gamma}\coth\sqrt{\gamma}l_\ab
  \right]^{-1}
 \right)_{\alpha\alpha}
  = \frac{ (\mathcal{M}_1^{-1})_{\alpha\alpha} }
         { 1 + (\mathcal{M}_1^{-1})_{\alpha\alpha}\:\sqrt{\gamma}\coth\sqrt{\gamma}l_\ab}
\end{equation}
a little bit of algebra gives~:
\begin{eqnarray}
  \label{reldet}
  \det\mathcal{M} &=&\det\mathcal{M}_1 \det\mathcal{M}_2 \nonumber\\
   &&\times
  \left\{
    1 
    + \sqrt\gamma\coth\sqrt{\gamma}l_\ab  
    \left[ 
      (\mathcal{M}_1^{-1})_{\alpha\alpha} + (\mathcal{M}_2^{-1})_{\beta\beta}
    \right]
    + \gamma (\mathcal{M}_1^{-1})_{\alpha\alpha}
             (\mathcal{M}_2^{-1})_{\beta\beta}
  \right\}
\end{eqnarray}
In the limit $\lambda_\alpha\to\infty$, corresponding to Dirichlet boundary
condition, the determinant behaves linearly with $\lambda_\alpha$, therefore
we define the spectral determinant with Dirichlet 
boundary condition at vertex~$\alpha$ and $\beta$ as ~:
\begin{eqnarray}
  S_1^\mathrm{Dir}(\gamma) =
    \lim_{\lambda_\alpha\to\infty} 
    \frac{S_1^{\lambda_\alpha}(\gamma)}{\lambda_\alpha}
  \:,\hspace{1cm}
  S_2^\mathrm{Dir}(\gamma) =
    \lim_{\lambda_\beta\to\infty} 
    \frac{S_2^{\lambda_\beta}(\gamma)}{\lambda_\beta}
\end{eqnarray}
The Dirichlet-determinant is computed by eliminating in $\det\mathcal{M}_1$
the column and the line corresponding to vertex $\alpha$. Therefore
$(\mathcal{M}_1^{-1})_{\alpha\alpha}=S_{1}^{\mathrm{Dir}}/S_{1}$. Finally,
using eqs.~(\ref{spedet},\ref{reldet}), we obtain for the spectral determinant
of the graph of figure~\ref{fig:attachgraphs}.b~:
\begin{eqnarray}
\label{RES0}
  \sqrt{\gamma}\, S_{\widetilde{\mathcal{G}}}(\gamma) = 
  &\cosh\sqrt{\gamma}l_\ab& 
    \left[ 
         S_{1}(\gamma)\,\sqrt{\gamma}S_{2}^{\mathrm{Dir}}(\gamma)
      + \sqrt{\gamma}S_{1}^{\mathrm{Dir}}(\gamma)\,S_{2}(\gamma) 
    \right] \nonumber\\[0.15cm]
  + &\sinh\sqrt{\gamma}l_\ab&
    \left[ 
       S_{1}(\gamma)\,S_{2}(\gamma) 
      +\sqrt{\gamma}S_{1}^{\mathrm{Dir}}(\gamma)\,
       \sqrt{\gamma}S_{2}^{\mathrm{Dir}}(\gamma)
    \right]
\end{eqnarray}

At this stage it is interesting to discuss the simple case of a graph with
Dirichlet boundary at vertices $\alpha$ and $\beta$. We can take the limit
$\lambda_\alpha,\,\lambda_\beta\to\infty$, which corresponds to the
subsitution $S_1\to\lambda_\alpha{}S_1^\mathrm{Dir}$ and 
$S_2\to\lambda_\beta{}S_2^\mathrm{Dir}$.
We obtain the expected result
$
\lim_{\lambda_\alpha,\,\lambda_\beta\to\infty}
\frac{S_{\widetilde{\mathcal{G}}}(\gamma)}{\lambda_\alpha\lambda_\beta}
=\frac{\sinh\sqrt{\gamma}l_\ab}{\sqrt{\gamma}}
S_{1}^{\mathrm{Dir}}(\gamma)S_{2}^{\mathrm{Dir}}(\gamma)
$
equivalent to
$
\mathrm{Spec}(-\Delta;\widetilde{\mathcal{G}})
=
\mathrm{Spec}(-\Delta;\mathcal{G}_1)
\cup
\mathrm{Spec}(-\Delta;\mathcal{G}_2)
\cup
\{(\frac{n\pi}{l_\ab})^2;\,n\in\NN^*\}$.

\vspace{0.25cm}

The last step of the graph attachment consists to take the limit $l_\ab\to0$
(figure~\ref{fig:attachgraphs}.c) we obtain~:
\begin{equation}
\label{RESULT}
  S_\mathcal{G}(\gamma) 
     = S_1(\gamma)\,             S_2^\mathrm{Dir}(\gamma)
     + S_1^\mathrm{Dir}(\gamma)\,S_2(\gamma) 
\end{equation}
which is the central result of the present article.

\vspace{0.25cm}

\noindent
{\it Example~: Ring attached to a wire (figure~\ref{fig:graphexample}).--}
If we consider for $\mathcal{G}_1$ a ring of perimeter $L$ pierced by a flux
$\theta$ (corresponding to $A(x)=\theta/L$), we have~:
$S_\mathrm{ring}=2(\cosh\sqrt{\gamma}L-\cos\theta)$ and
$S^\mathrm{Dir}_\mathrm{ring}=\frac{\sinh\sqrt{\gamma}L}{\sqrt{\gamma}}$. The
graph $\mathcal{G}_2$ is a wire of length $b$ with Neumann boundary at its ends
(for a vertex $\alpha$ of coordination number $m_\alpha=1$ the case $\lambda_\alpha=0$
coincides with Neumann boundary condition)~:
$S_\mathrm{wire}^\mathrm{both\:Neu}=\sqrt{\gamma}\sinh\sqrt{\gamma}b$ and
$S_\mathrm{wire}^\mathrm{Neu/Dir}=\cosh\sqrt{\gamma}b$ . Therefore, we recover
the simple result~:
$
S(\gamma)  = \sinh\sqrt{\gamma}b\,\sinh\sqrt{\gamma}L
+ 2 \cosh\sqrt{\gamma}b\,(\cosh\sqrt{\gamma}L-\cos\theta)
$
obtained directly from~(\ref{spedet}) in Ref.~\cite{TexMon05}.


\vspace{0.25cm}

\noindent{\it Eq.~(\ref{RES0}) is contained in eq.~(\ref{RESULT}).--}
The result (\ref{RESULT}) has appeared as a limit of eq.~(\ref{RES0}),
therefore it seems at first sight a particular case of this latter equation.
We show now that eq.~(\ref{RES0}) can in fact be recovered from (\ref{RESULT}). For
that purpose we proceed in two steps.
First 
we attach a wire of length $b$ to the graph $\mathcal{G}_1$. The graph formed
is denoted $\mathcal{G}_\mathrm{inter}$ and the corresponding spectral
determinants $S_\mathrm{inter}$ and
$S_\mathrm{inter}^{\mathrm{Dir}}$, depending on the nature of the boundary
condition at the end of the wire. Spectral determinants of the wire for the
three different boundary conditions, $S_\mathrm{wire}^\mathrm{both\:Neu}$,
$S_\mathrm{wire}^\mathrm{Neu/Dir}$ and
$S_\mathrm{wire}^\mathrm{both\:Dir}\equiv{S}^\mathrm{Dir}_\mathrm{ring}$, were
given above. Therefore, from eq.~(\ref{RESULT}) we obtain~:
\begin{eqnarray}
  S_\mathrm{inter} &=&
    S_{1}\,S_\mathrm{wire}^\mathrm{Neu/Dir}
  + S_{1}^{\mathrm{Dir}}\,S_\mathrm{wire}^\mathrm{both\:Neu}
  = 
  S_{1}\, \cosh\sqrt{\gamma}b
+ S_{1}^{\mathrm{Dir}}\, \sqrt{\gamma}\sinh\sqrt{\gamma}b \\
  S_\mathrm{inter}^{\mathrm{Dir}} &=& 
    S_{1}\,S_\mathrm{wire}^\mathrm{both\:Dir}
  + S_{1}^{\mathrm{Dir}}\,S_\mathrm{wire}^\mathrm{Neu/Dir}
  \hspace{0.1cm}=
   S_{1}\, \frac{\sinh\sqrt{\gamma}b}{\sqrt{\gamma}}
 + S_{1}^{\mathrm{Dir}}\, \cosh\sqrt{\gamma}b
\end{eqnarray}
In a second step we attach the graph $\mathcal{G}_2$ to the end of the wire of
$\mathcal{G}_\mathrm{inter}$.
We use again eq.~(\ref{RESULT}) from which it follows that
$
S_{\widetilde{\mathcal{G}}}
=S_\mathrm{inter}S_{2}^{\mathrm{Dir}}
+S_\mathrm{inter}^{\mathrm{Dir}}S_{2}
$,
that precisely coincides with eq.~(\ref{RES0}).

\vspace{0.25cm}

\noindent{\bf Attachment of two graphs for Schr\"odinger operator.--}
Let us consider the spectral determinant for the Schr\"odinger operator (Hill
operator) $S_\mathcal{G}(\gamma)=\det(\gamma-\Delta+V(x))$ with the same
(continuous) boundary conditions as above. Let us first discuss how
(\ref{spedet},\ref{defmatM}) are modified. $V_\ab(x_\ab)$, with
$x_\ab\in[0,l_\ab]$, is the component of the scalar potential $V(x)$ on the
bond. An important ingredient is the solution $f_\ab(x_\ab)$ of the
differential equation
$[\gamma-\frac{\D^2}{\D{x_\ab^2}}+V_\ab(x_\ab)]f_\ab(x_\ab)=0$ on the interval
$[0,l_\ab]$, satisfying $f_\ab(0)=1$ and $f_\ab(l_\ab)=0$. A second
independent solution of the differential equation is
$f_\ba(x_\ba)=f_\ba(l_\ab-x_\ab)$ (one should not make a confusion~: despite
we use the same notation, the $2B$ functions $f_\ab$ are not the $B$ components of
a scalar function). If $V_\ab(x_\ab)=0$ we have obviously
$f_\ab(x)=\frac{\sinh\sqrt\gamma(l_\ab-x)}{\sinh\sqrt\gamma{}l_\ab}$. It was
shown in Ref.~\cite{Des00}\footnote{
  The notations used here are slightly different from those of Ref.~\cite{Des00}.
  They coincide with those of Refs.~\cite{TexDeg03,ComDesTex05}. 
}
that eqs.~(\ref{spedet},\ref{defmatM}) are generalized by performing the 
substitution
$\sqrt\gamma\coth\sqrt{\gamma}l_\ab\to-f_\ab'(0)$ and 
$\frac{\sqrt\gamma}{\sinh\sqrt{\gamma}l_\ab}\to-f_\ab'(l_\ab)$.
The matrix $\mathcal{M}$ becomes
$
\mathcal{M}_\ab = 
\delta_\ab[\lambda_\alpha - \sum_\mu a_{\alpha\mu}f_{\alpha\mu}'(0)]
+a_\ab f_\ab'(l_\ab)\,\EXP{-\I\theta_\ab}
$
and the spectral determinant takes the form\footnote{
  \label{Afootnote}
  The fact that $S_\mathcal{G}(\gamma)\propto\det\mathcal{M}$ has already been
  demonstrated in Ref.~\cite{Exn97b}, however the remaining factor has been
  obtained in Ref.~\cite{Des00} by construction of the resolvent in the graph.
  Note that the Dirichlet determinant $[\prod_{(\ab)}f_\ab'(l_\ab)]^{-1}$ may play 
  a role in order to determine the full spectrum. A trivial example is
  the wire (with $V(x)=0$) for which $\det\mathcal{M}=\gamma$, that does not 
  determine the spectrum. Another example is
  studied in detail in section~12 of Ref.~\cite{AkkComDesMonTex00}.
  The $\gamma$-dependent factor $[\prod_{(\ab)}f_\ab'(l_\ab)]^{-1}$  is also
  important from a physical point of view since
  $\drond{}{\gamma}\ln{S(\gamma)}$ (for $V(x)=0$) has been shown to be 
  related to several physical quantities~\cite{PasMon99,AkkComDesMonTex00,ComDesTex05}.
}
$S_\mathcal{G}(\gamma)=[\prod_{(\ab)}f_\ab'(l_\ab)]^{-1}\det\mathcal{M}$.

We consider two graphs $\mathcal{G}_1$ and $\mathcal{G}_2$ on which lives a
scalar potential $V(x)$. If these two graphs are attached by a bond $(\ab)$
where the potential vanishes [$V(x)\neq0$ for
$x\in\mathcal{G}_1\cup\mathcal{G}_2$ and $V(x)=0$ for $x\in(\ab)$], the
structure (\ref{structureM}) still holds. Therefore all results derived above
are still valid, and in particular eqs.~(\ref{RES0},\ref{RESULT}) and
also~(\ref{RESn}).

\vspace{0.25cm}

\mathversion{bold}
\noindent{\bf Attachment of $n$ graphs.--}
\mathversion{normal} We consider a graph $\mathcal{G}$ obtained by attachment
of $n$ graphs at the same point (figure~\ref{fig:attachgraphs}.d). It is now
easy to generalize (\ref{RESULT}) in order to describe this situation. We
start from (\ref{RESULT})~:
$S_\mathcal{G}=S_1\,S^\mathrm{Dir}_{2+\cdots+n}+S^\mathrm{Dir}_1\,S_{2+\cdots+n}$
and use
$S^\mathrm{Dir}_{2+\cdots+n}=S^\mathrm{Dir}_2\cdots{S}^\mathrm{Dir}_n$.
Proceeding by recurrence, we end with~:
\begin{equation}
  \label{RESn}
  S_\mathcal{G} = 
  \sum_{k=1}^n 
  \underbrace{S_1^\mathrm{Dir}\cdots S_{k-1}^\mathrm{Dir}}_{\mathrm{Dirichlet}}\,
  S_{k}\,
  \underbrace{S_{k+1}^\mathrm{Dir}\cdots S_n^\mathrm{Dir}}_{\mathrm{Dirichlet}}
\:.
\end{equation}

\vspace{0.25cm}

\begin{figure}[!ht]
  \centering
  \diagram{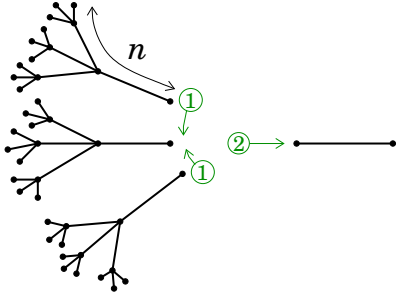}{0.8}{-1cm}
  \caption{\it Cayley tree of coordination number~$z$ (here $4$) and depth $n$
    (here $3$ before attachement and $4$ after).}
  \label{fig:cayley}
\end{figure}

\noindent{\bf Cayley tree.--}
We can use (\ref{RESn}) to study the case of a Cayley tree of coordination number~$z$.
We denote $S_n$ the spectral determinant of a Cayley tree of
depth $n$ with $\lambda_\alpha=0\hspace{0.3cm}\forall\,\alpha$. The spectral
determinant for the similar graph with Dirichlet boundary at one of its end is
denoted $S_n^\mathrm{Dir}$. We proceed in two steps represented on
figure~\ref{fig:cayley}~: first we attach $z-1$ such trees together, using
(\ref{RESn}). Then we attach a wire of length $b$ by using (\ref{RESULT}). We
find~:
\begin{eqnarray}
  S_{n+1} &=& 
  \left[ 
    (z-1)\, S_n                          \: \cosh\sqrt\gamma b
    +       \sqrt\gamma S_n^\mathrm{Dir} \: \sinh\sqrt\gamma b
  \right]
  \left(S_n^\mathrm{Dir}\right)^{z-2}\\ 
  \sqrt\gamma S_{n+1}^\mathrm{Dir} &=&   
  \left[ 
    (z-1)\, S_n                          \: \sinh\sqrt\gamma b
    +       \sqrt\gamma S_n^\mathrm{Dir} \: \cosh\sqrt\gamma b
  \right]
  \left(S_n^\mathrm{Dir}\right)^{z-2}
\end{eqnarray}
with $S_1=S_\mathrm{wire}^\mathrm{both\:Neu}$ and
$S_1^\mathrm{Dir}=S_\mathrm{wire}^\mathrm{Neu/Dir}$. Note that in the case
$z=2$ the recurrence is trivially solved and give the spectral determinants
for a wire of length~$nb$.

\vspace{0.25cm}

\noindent{\bf Conclusion.--}
Let us come back to the initial question of the paper. Eq.~(\ref{RESULT}) seems at
first sight to involve only spectral information on graphs $\mathcal{G}_1$,
$\mathcal{G}_2$ and $\mathcal{G}$, however we mentioned above that, in
eq.~(\ref{spedet}), the $\gamma$-independent prefactor is not {\it a priori}
fixed by spectral information.
Therefore we can only provide here a partial answer to the initial question~:
given the spectral determinants of two metric graphs $\mathcal{G}_1$ and
$\mathcal{G}_2$, defined by (\ref{spedet}), we can determine the spectrum of
the graph $\mathcal{G}$ formed by attaching the two graphs at a vertex.
An interesting development would be to provide a relation similar to
(\ref{RESULT}) when the spectral determinant {\it and} its
$\gamma$-independent prefactor are constructed from the spectrum only. In the
case of $\zeta$-regularization of Ref.~\cite{Fri06}, the relation between
$\zeta$-regularized determinant and (\ref{spedet}), mentioned in a footnote
above, suggests that the relation for $\zeta$-regularized determinants
analogous to (\ref{RESULT}) also involves some information on the coordination
numbers of vertices.

The choice of continuous boundary conditions was an important hypothesis in
order to derive eqs.~(\ref{RESULT},\ref{RESn}). Another simple choice of
boundary conditions, assuming continuity of the derivative of the field at the
vertices, is examined in the appendix. This leads to a relation with a
similar structure, eq.~(\ref{RESnNeumann}). A question would be to generalize
the results (\ref{RESn},\ref{RESnNeumann}) to the case of general boundary
conditions. This would require to formulate the problem with matrices coupling
arcs~\cite{Des01} since in absence of continuity of the field or its
derivative, one cannot introduce anymore vertex variables.

An interesting development would be to generalize (\ref{RESULT}) to other
attachment procedures (graphs attached at more than one vertex). For that
purpose, a helpful starting point may be the scattering interpretation of
equation (\ref{RESULT}). If a graph is connected to an infinite wire, its
spectrum is continuous and we can consider the scattering problem. A plane
wave $\EXP{-\I{kx}}$ of energy $E=-\gamma=k^2$ sent from the infinite lead is
reflected by the graph with a phase shift $\EXP{\I{kx}+\I\delta(k^2)}$ given
by $\cotg[\delta(E)/2]=-\sqrt{E}\frac{S^\mathrm{Dir}(-E)}{S(-E)}$, as shown in
Ref.~\cite{AkkComDesMonTex00} (eq.~(117)), where $S^\mathrm{Dir}(-E)$
corresponds to Dirichlet boundary condition at the vertex where infinite wire
is attached. We can associate to the two graphs $\mathcal{G}_1$ and
$\mathcal{G}_2$ two such phase shifts $\delta_1(E)$ and $\delta_2(E)$. The
spectrum of the graph $\mathcal{G}$ obtained by attachement of $\mathcal{G}_1$
and $\mathcal{G}_2$ is given by the Bohr-Sommerfeld quantization condition
$\delta_1(E_n)+\delta_2(E_n)=2n\pi$, that rewrites
$\cotg[\delta_1/2]+\cotg[\delta_2/2]=0$. Since the spectral determinant
vanishes on the spectrum, $S(-E_n)=0$, this shows that
$S\propto\frac{S_1^{\mathrm{Dir}}}{S_1}+\frac{S_2^{\mathrm{Dir}}}{S_{2}}$
(note however that this argument misses a factor function of the energy~; see
the footnote~\ref{Afootnote}). The scattering problem has been studied for
graphs with an arbitrary number of contacts (infinite leads)~; in particular
expressions of the scattering matrix of a graph with $L$ infinite leads is
available in Ref.~\cite{AvrSad91} (for $V(x)=0$) and \cite{TexMon01} (for
$V(x)\neq0$). 
It must also be pointed that the question of graph attachment has been studied
in Ref.~\cite{KosSch01} and in particular how to construct the scattering
matrix of a graph in terms of subgraphs scattering matrices. All these results
on scattering theory in graphs might help the construction of the spectral
determinant of two graphs attached by $L>1$ vertices.

\vspace{0.25cm}

\noindent{\bf Acknowledgments.--}
I thank Y.~Colin~de~Verdi\`ere and Alain Comtet for interesting discussions.

\vspace{0.25cm}

\noindent{\bf Appendix~: derivative continous at the vertices.--}
The boundary conditions discussed in this article ($\varphi(x)$ continuous and
$\sum_\beta{a}_\ab\varphi_\ab'(0)=\lambda_\alpha\,\varphi_\alpha$) can be
interpreted as the introduction of a $\delta$-potential at the vertex. They
are denoted ``$\delta$-coupling'' in Ref.~\cite{Exn97b}, where
``$\delta'$-coupling'' are also introduced. These latter correspond to
continuity of the derivative~:
$\varphi'_\ab(0)=\varphi'_\alpha\:\forall\:\beta$ neighbour of $\alpha$ and
$\sum_\beta{a}_\ab\varphi_\ab(0)=\mu_\alpha\,\varphi'_\alpha$ (the limit
$\mu_\alpha\to\infty$ corresponds to Neumann boundary condition
$\varphi'_\alpha=0$). The results of the present article are easily
generalized to the case of $\delta'$-couplings.

\noindent{\it Spectral determinant.--}
The spectral determinant now involves the solution of the differential equation
$[\gamma-\frac{\D^2}{\D{x_\ab^2}}+V_\ab(x_\ab)]g_\ab(x_\ab)=0$ on the interval
$[0,l_\ab]$, satisfying $g'_\ab(0)=1$ and $g'_\ab(l_\ab)=0$. The spectral
determinant is given by
$S_\mathcal{G}(\gamma)=[\prod_{(\ab)}g_\ab(l_\ab)]^{-1}\det\mathcal{N}$
with 
$
\mathcal{N}_\ab = 
\delta_\ab[\mu_\alpha - \sum_\nu a_{\alpha\nu}g_{\alpha\nu}(0)]
-a_\ab g_\ab(l_\ab)\,\EXP{-\I\theta_\ab}
$.
Note that $[\prod_{(\ab)}g_\ab(l_\ab)]^{-1}$ corresponds to the Neumann
determinant~: disconnected wires with Neumann boundary conditions 
$\mu_\alpha\to\infty\:\forall\:\alpha$).
In the absence of a potential, $V(x)=0$~:
\begin{equation}
  \label{defmatN}
  \mathcal{N}_\ab = 
  \delta_\ab
  \left(\mu_\alpha +\frac1{\sqrt{\gamma}} \sum_\nu a_{\alpha\nu}
  \coth\sqrt{\gamma}l_{\alpha\nu}\right)
  +a_\ab \frac{\EXP{-\I\theta_\ab}}{\sqrt\gamma\sinh\sqrt{\gamma}l_\ab}
  \:.
\end{equation}
and
$
S_\mathcal{G}(\gamma)=(\prod_{(\ab)}\sqrt\gamma\sinh\sqrt{\gamma}l_\ab)\:\det\mathcal{N}
$.

\noindent{\it Graph attachment.--}
We consider $n$ Graphs charaterized by spectral determinants $S_k$ for
$k=1,\cdots,n$. We introduce the notation
$S_k^\mathrm{Neu}=\lim_{\mu_\alpha\to\infty}\frac{S_k}{\mu_\alpha}$, where
$\alpha$ is the vertex of attachment of the $n$ graphs
(figure~\ref{fig:attachgraphs}.d). Since spectral determinants for continuous
boundary conditions and continuous derivative have similar structures, the
results obtained in this article are easily generalized. In particular the
result (\ref{RESn}), from which other results have been derived, becomes for
$\delta'$-couplings
\begin{equation}
  \label{RESnNeumann}
  S_\mathcal{G} = 
  \sum_{k=1}^n 
  \underbrace{S_1^\mathrm{Neu}\cdots S_{k-1}^\mathrm{Neu}}_{\mathrm{Neumann}}\,
  S_{k}\,
  \underbrace{S_{k+1}^\mathrm{Neu}\cdots S_n^\mathrm{Neu}}_{\mathrm{Neumann}}
\:.
\end{equation}



\end{document}